\documentstyle[prl,aps,twocolumn,epsf]{revtex}
\begin{document}
\title{Spontaneous emission of phonons by coupled quantum dots} 
\author{T. Brandes, B. Kramer} 
\address{University of Hamburg,
1. Inst. Theor. Physik, Jungiusstr. 9,
D-20355 Hamburg, Germany}
\draft
\date{\today}
\maketitle
\begin{abstract}
We find an interference effect for electron--phonon interactions in
coupled semiconductor quantum dots that can dominate the nonlinear transport properties
even for temperatures close to zero.
The intradot electron tunnel process leads to a `shake up'
of the phonon system 
and is dominated by a double-slit-like interference effect
of spontaneously emitted phonons. 
The effect is closely related to subradiance of photons in a laser--trapped two--ion system
and explains the oscillations in the nonlinear current-voltage characteristics of
coupled dots observed recently. 
\end{abstract} 

Spontaneous emission is one of the fundamental concepts of quantum mechanics that can be traced back to
such early works as that of Albert Einstein \cite{Ein17}. An excited state of a single
atom decays exponentially due to the coupling to photons.
In a system of {\em two} atoms interacting via the common photon field, the decay splits into
a sub- and a superradiant channel. This effect is a precursor of the famous Dicke superradiance
phenomenon \cite{Dic54} and was verified experimentally in the spontaneous emission of photons
from two trapped ions only three years ago \cite{DeVB96}.

Recently, in a completely
different physical system, the emission of {\em phonons}
from two artificial atoms has been observed \cite{Fujetal98}. Here,
the coupling to the phonon degrees of freedom turned out to dominate 
the non--linear electron transport through semiconductor double quantum dots
even at $mK$ temperatures.

Double quantum dots are well-defined artificial systems for the study of
interaction  \cite{ddinteraction} and coherent time-dependent \cite{ddcoherent}
effects.
Here, we propose a theory for a new non--linear transport effect in
double quantum dots which corresponds to the Dicke effect, i.e. the collective decay of real atoms.
In our theory, the tunneling of single electrons through coupled artificial atoms is renormalized
by the interaction with piezoelectric acoustic phonons and leads to an orthogonality
catastrophe of the phonon bath if an electron tunnels
between the dots. This `boson shake up' effect \cite{Giretal90,IN91}
is determined by an effective density of states $\rho(\omega)$ of the phonon modes ${\bf Q}$ that
couple to the tunnel process. These  
interfere like in a double slit experiment when interacting with the
electron densities in the two dots. As a result,  $\rho(\omega)$
shows oscillations on a scale $\omega_d:=c_s/d$, where $c_s$ is the speed of sound and $d$ the
distance between the centers of the two dots. It turns out that the non--linear current peak as a function of the 
difference $\varepsilon$ between the two relevant many-particle energies
is determined by the shape of $\rho(\omega=\varepsilon/\hbar)$.
Furthermore, this quantity is analogous
to the rate for emission of subradiant {\em photons} from two laser--trapped
ions \cite{DeVB96}, when $c_s$ is replaced by the speed of light and $d$ by the distance of the ions.
Thus, both phenomena are physically closely related. This provides  
the microscopic mechanism for the oscillations observed recently in a double
dot current spectrum \cite{Fujetal98}. We predict that future experiments with artificial atoms
can exploit this analogy to real atoms in more detail.
In particular, coherent effects such as  superradiance \cite{Benedict} can 
be  manipulated by gate--voltages and external leads in such systems \cite{BIS98}.

In our model, we consider a double quantum dot consisting of a left and a right dot (L and R,
respectively)
which are connected via a tunnel barrier. Each dot is connected to an
electron reservoir in thermal equilibrium
with chemical potentials $\mu_L$ (source) and $\mu_R<\mu_L$ (drain).
For the physical phenomena we are interested in, it is sufficient
\cite{SN96} to restrict the dot Hilbert space to the three states
$
|0\rangle =|N_L,N_R\rangle$,
$|L\rangle =|N_L+1,N_R\rangle$, and 
$|R\rangle =|N_L,N_R+1\rangle$,
which correspond to many-particle ground states
with $N_L(+1)$ electrons in the left and $N_R(+1)$ electrons in the right dot.
The corresponding ground state energies $\varepsilon_L$ of $|L\rangle$ and
$\varepsilon_R$ of $|R\rangle$ are in the window 
$\mu_L>\varepsilon_L,\varepsilon_R>\mu_R$.
We assume that the Coulomb charging energy $U_c$ is
the largest energy scale, and it is not possible to
charge the double dot with more than one additional electron.
In \cite{Fujetal98}, no enhanced tunnel current was
observed for $\varepsilon:=\varepsilon_L-\varepsilon_R<0$ at low temperatures so that excited
many-body states play no role. In particular, $U_c\sim 1$ meV was one order of magnitude larger
than the external source drain voltage $V_{SD}$. This situation has to be contrasted
with the regime $V_{SD}\gtrsim U_c$ (absence of blockade effects, \cite{RaikhAsenov}).

We define operators
$n_L=|L\rangle \langle L|$, $n_R=|R\rangle \langle
R|$, $p=|L\rangle \langle R|$,
$s_L=|0\rangle \langle L|$, $s_R=|0\rangle \langle R|$, and 
the total system Hamiltonian $H$ as the sum of the dot, the phonon, the reservoir
and the electron-phonon interaction 
\begin{eqnarray}\label{hamiltonian}
  H&=&H_0'+ H_T+H_V + H_{\alpha\beta}\nonumber\\
  H_0'&=&\varepsilon_Ln_L+\varepsilon_Rn_R+H_{p}+H_{res}\nonumber\\
  H_T&=&T_c(p+p^{\dagger}),\quad  H_p= \sum_{\bf{Q}}\omega_{\bf{Q}}
  a^{\dagger}_{\bf{Q}}a_{\bf{Q}}\nonumber\\
 H_{\alpha\beta}&=&\sum_{\bf{Q}}\left(
  \alpha_{\bf{Q}}n_L+\beta_{\bf{Q}}n_R\right)
  \left(a_{-\bf{Q}}+a^{\dagger}_{\bf{Q}}\right)\nonumber\\
  H_V&=&\sum_{\bf k}\left(V_{\bf k}c_{\bf k}^{\dagger}s_L+W_{\bf k}d_{\bf k}^{\dagger}s_R +
  c.c.\right)\nonumber\\
H_{res}&=&\sum_{\bf k}\varepsilon^L_{\bf k}c_{\bf k}^{\dagger}c_{\bf k}^{\phantom{\dagger}}
+\sum_{\bf k}\varepsilon^R_{\bf k}d_{\bf k}^{\dagger}d_{\bf k}^{\phantom{\dagger}}.
\end{eqnarray}
Here, the tunneling between left and right dots is described by
a single tunnel matrix element $T_c$.
In the standard tunnel Hamiltonian $H_V$, 
$V_{\bf k}$ and $W_{\bf k}$ couple the dot to a continuum of channels ${\bf k}$
of the left and right electron reservoir $H_{res}$.
The spin of the electron does not play any role here and is suppressed.
The term $H_p$ describes the lattice vibrations in harmonic approximation; 
the creation operator for a phonon of mode ${\bf Q}$ is $a^{\dagger}_{{\bf Q}}$.
The electron--phonon matrix elements are defined by 
$\alpha_{\bf Q}:=\lambda_{\bf Q}\langle L|e^{i{\bf Qr}}|L\rangle$ and 
$\beta_{\bf Q}:=\lambda_{\bf Q}\langle R|e^{i{\bf Qr}}|R\rangle$,
where $\lambda_{\bf Q}$ is the matrix element for the interaction of 2DEG electrons and
phonons.
We have already neglected the non-diagonal term of the electron-phonon interaction
which contains matrix elements $\gamma_{\bf Q}:=\lambda_{\bf Q}\langle L|e^{i{\bf Qr}}|R\rangle$.
We have checked in a seperate master equation calculation that such terms modify only weakly the
tunnel current and do not lead to the oscillatory phenomena observed in \cite{Fujetal98}, which
is due to the non-perturbative shake-up process that we describe in the following.

Suppose an electron tunnels between two regions of space ($L$ and $R$) and interacts with a phonon field. With the interaction
of the form $H_{\alpha\beta}$, Eq.(\ref{hamiltonian}), the electron--phonon coupling locally changes the energy of the
electron, depending on whether it is in $L$ or in $R$. During  tunneling,
its wave function experiences an additional a phase shift
$e^{i\phi}$ due to this coupling. This phase shift is zero if the coupling is identical in both regions, i.e. if $\alpha_{\bf Q}=
\beta_{\bf Q}$. If,
as we show below, $ \alpha_{\bf Q}\ne\beta_{\bf Q}$, the phase shift depends on the state of the phonon field.
From the point of view of the phonons,
initial (before the tunneling) and final (after tunneling) phonon states are no longer the same. 
From the point of view of the electron, the effective tunnel amplitude changes non--trivially.
In particular, it becomes time-dependent.

We introduce now a unitary polaron transformation \cite{Mahan}
of the Hamiltonian that naturally leads to the phase factors mentioned above.
For any operator $O$, we define $\overline{O}:=e^SOe^{-S}$,
$S:=n_LA+n_RB$ with $A:=$ $\sum_{\bf{Q}}(1/\omega_{\bf Q})$ $
(\alpha_{\bf{Q}} a^{\dagger}_{\bf{Q}}-\alpha_{-\bf{Q}}a_{\bf{Q}}^{\phantom{\dagger}})$ and
$B:=$ $\sum_{\bf{Q}}(1/\omega_{\bf Q})$ $
(\beta_{\bf{Q}} a^{\dagger}_{\bf{Q}}-\beta_{-\bf{Q}}a_{\bf{Q}}^{\phantom{\dagger}})$. 
This leads to renormalized energies $
\overline{\varepsilon_L}=\varepsilon_L-
\sum_{\bf{Q}}|\alpha_{\bf{Q}}|^2/\omega_{\bf{Q}}$ and
$\overline{\varepsilon_R}=\varepsilon_R-
\sum_{\bf{Q}}|\beta_{\bf{Q}}|^2/\omega_{\bf{Q}}$, and
a renormalized intra--dot tunneling Hamiltonian
$\overline{H}_T=T_c(pX+p^{\dagger}X^{\dagger})$.
Here, the phase operator
$X=\prod_{\bf{Q}}D_{\bf{Q}}\left((\alpha_{\bf{Q}}-\beta_{\bf{Q}})/\omega_{\bf{Q}}\right)$
is the product of {\em unitary displacement operators}
$ D_{\bf{Q}}(z):=\exp({za^{\dagger}_{\bf{Q}}-z^*a_{\bf{Q}}^{\phantom{\dagger}}})$,
where the operation of $D_{\bf{Q}}(z)$ on the boson vacuum
creates a coherent state of the boson field  mode ${\bf Q}$.
The factors $X$ and $X^{\dagger}$ in the tunnel Hamiltonian $\overline{H}_T$
drastically change the transport properties of the double dot.

We assume that the coupling to the
left and right electron reservoirs is weak such that a standard Born and Markov approximation 
holds.
In contrast to this, we calculate
to all orders of the  {\em intra}dot tunneling $T_c$
and the electron-phonon coupling because
the renormalization  of the tunneling by the
phase factors $X$ is a non-perturbative effect.
From the Liouville
equation for the total density matrix of the system, one obtains the equations of motion
(cp. \cite{SN96,GP96})
\begin{eqnarray}\label{eom1}
\frac{d}{dt}\langle n_L\rangle_t&=&-iT_c\left(
\langle p \rangle_{t}-\langle p^{\dagger}\rangle_{t} \right) +
2\Gamma_L(1-\langle n_L\rangle_{t} - \langle n_R\rangle_{t})\nonumber\\
\frac{d}{dt}\langle
n_R\rangle_t&=&iT_c\left(\langle
p\rangle_{t}-\langle p^{\dagger}\rangle_{t} \right) 
-2\Gamma_R\langle n_R\rangle_{t} \\
\langle
p\rangle_t&=&-\Gamma_R\int_0^tdt' e^{i\varepsilon(t-t')}  \langle
X_t^{\phantom{\dagger}}X_{t'}^{\dagger}\tilde{p}(t')\rangle_{t'}\nonumber\\
-iT_c\int_0^t&dt'&e^{i\varepsilon(t-t')}\left( \langle
n_LX_t^{\phantom{\dagger}}X_{t'}^{\dagger}\rangle_{t'}- \langle
n_RX_{t'}^{\dagger}X_t^{\phantom{\dagger}}\rangle_{t'}\right)\nonumber, 
\end{eqnarray}
where 
$\Gamma_L:=$ $2\pi$ $\sum_{\bf k}V_{\bf k}^2$ $\delta(\overline{\varepsilon}_L-\varepsilon^L_{\bf k})$, 
$\Gamma_R:=$ $2\pi$ $\sum_{\bf k}W_{\bf k}^2$ $\delta(\overline{\varepsilon}_R-\varepsilon^R_{\bf k})$,
and the chemical potential $\mu_L$ ($\mu_R$) of the left (right)  electron reservoir is assumed
to be far above (below) $\overline{\varepsilon}_L$ ($\overline{\varepsilon}_R$)
so that no electrons can tunnel from the left dot into the left reservoir, or from the right reservoir into the right dot.
Furthermore, $\varepsilon=$ $\overline{\varepsilon}_L-\overline{\varepsilon}_R$, where we neglect the difference
in the energy renormalization in both dots, $\tilde{p}(t)=pe^{i\varepsilon t}X_t$, and $X_t$ denotes the
time-evolution of $X$ with $H_p$.

We assume the phonon system to be in thermal
equilibrium at all times. We use a decoupling of the reduced density matrix $\tilde{\rho}(t')$ of the dot,
$\tilde{\rho}(t')\approx
\rho_{ph}^0{\rm Tr}_{ph}\{\tilde{\rho}(t')\}$. We introduce
the Laplace transform
\begin{equation}\label{equ}
C_{\varepsilon}(z):=\int_0^{\infty}dt e^{-(z-i\varepsilon) t}
\langle X_t^{\phantom{\dagger}}X_{0}^{\dagger}\rangle_0
\end{equation}
and $n_L(z):=\int_0^{\infty}dt e^{-zt} \langle n_L\rangle_{t}$ etc. for $z>0$.
By transforming Eq.(\ref{eom1}) into $z$-space,
the expectation value of the current operator $\hat{I}:=$  
$iT_c(p-p^{\dagger})$ (we set the electron charge $e=1$ for convenience), can 
be obtained in the stationary limit $t\to \infty$ from the
$1/z$-coefficient of the $I(z)$-expansion into a Laurent series for $z\to 0$. 
The result is 
\begin{eqnarray}\label{currentstat}
  & &\langle I \rangle_{t\to
  \infty}=T_c^2\frac{2\Re(C_{\varepsilon})+2\Gamma_R|C_{\varepsilon}|^2}
  {|1+\Gamma_RC_{\varepsilon}|^2+2T_c^2B_{\varepsilon}}\nonumber\\
B_{\varepsilon}&:=&
\Re\left\{(1+\Gamma_RC_{\varepsilon})\left[
 \frac{C_{-\varepsilon}}{2\Gamma_R}+\frac{C_{\varepsilon}^*}{2\Gamma_L}\left(1+\frac{\Gamma_L}{\Gamma_R}\right)\right]\right\}.
\end{eqnarray}
Here, 
$C_{\varepsilon}=C_{\varepsilon}(\delta \to 0)$.
One can verify that in the limit of no phonon coupling, $C_{\varepsilon}=i/\varepsilon$, thus obtaining
a simple Lorentzian curve for the stationary tunnel current as a function of 
$\varepsilon$ \cite{SN96}.
The appearance of $T_c$ in the denominator of
Eq.(\ref{currentstat}) indicates that this result is non--perturbative, i.e. valid to all orders
in $T_c$. The modification of this curve by the electron--phonon interaction is completely described by the
form of $C_{\varepsilon}$ which we will discuss now. 

For a harmonic phonon system in thermal equilibrium at 
temperature $T$, one finds the correlation function
$\langle X_t^{\phantom{\dagger}}X_{0}^{\dagger}\rangle_0$
$=$ $e^{-\Phi (t-t')}$
with
\begin{eqnarray}\label{oscillator}
\Phi (t)&:=&
\int_0^{\infty}d\omega \rho (\omega )\left\{(1-\cos \omega t)\coth
\frac{\hbar \omega}{2k_BT}+i\sin \omega t \right\}\nonumber\\
\rho (\omega)&=&\sum_{\bf{Q}}
\frac{|\alpha_{\bf{Q}}-\beta_{\bf{Q}}|^{2}}{\omega^2}\delta
(\omega -\omega _{\bf{Q}}).
\end{eqnarray}
The matrix elements $\alpha_{\bf Q}$ and $\beta_{\bf Q}$ are
$\alpha_{\bf Q}=\lambda_{\bf Q} \int d^3{\bf x}e^{i{\bf Qx}}\rho_L({\bf x})$,
$\beta_{\bf Q}=\lambda_{\bf Q} \int d^3{\bf x}e^{i{\bf Qx}}\rho_R({\bf x})$,
with the local electron densities $\rho_L({\bf x})$ and $\rho_R({\bf x})$ in the left and the
right dot, respectively. The exact form of both $\rho_L({\bf x})$ and $\rho_R({\bf x})$ depends on the
shape of the  dots and on the number of electrons $N_L$ and $N_R$. In the
stationary state for $t\to \infty$, both densities can be assumed to be
smooth functions of ${\bf x}$,
$\rho_L({\bf x})\approx \rho_e({\bf x-x}_L)$, $\quad \rho_R({\bf x})\approx \rho_e({\bf x-x}_R)$, 
where we have assumed that both left and right electron densities are described by the same  profile.
Here, $\rho_e({\bf x})$ is relatively sharply peaked around zero.
One obtains
$ \alpha_{\bf Q}=$  $\lambda_{\bf Q} e^{i{\bf Qr}_L}P({\bf Q})$,
$ \beta_{\bf Q} =$  $\lambda_{\bf Q} e^{i{\bf Qr}_R}P({\bf Q})$, where
$P({\bf Q})     =$  $\int d^3{\bf x}e^{i{\bf Qx}}\rho_e({\bf x})$.
Therefore,
$\beta_{\bf{Q}}= \alpha_{\bf{Q}}e^{i{\bf{Qd}}}$ 
with ${\bf{d}}={\bf{r}}_R-{\bf{r}}_L$.
The matrix elements $\alpha_{\bf Q}$ and $\beta_{\bf Q}$ thus coincide up to the phase
factor $e^{i{\bf{Qd}}}$. This phase factor
is the origin of the oscillations in the effective density of states $\rho(\omega)$
of phonon modes that couple to the tunnel process. We note that
the explicit form of $\alpha_{\bf Q}$ and $\beta_{\bf Q}$ can be
calculated exactly when the two states $|L\rangle$ and $|R\rangle$ are few
particle wave functions.

For sharp charge density profiles 
with Fourier transform  $P({\bf Q}) \to 1$,
we find
\begin{equation}\label{rho2}
  \rho(\omega)\approx \frac{g}{\omega}\left(1-\frac{\omega_d}{\omega}\sin\left(\frac{\omega}{\omega_d}\right)\right)
 e^{-\omega/\omega_c},\quad
  g=\frac{\lambda^2}{\pi^2\hbar^2c_s^3},
\end{equation}
where $\quad \omega_d:=c_s/d$, $d=|{\bf d}|$,
$c_s$ is the longitudinal speed of sound,
and a smooth exponential cutoff $\omega_c$
is an effective Debye frequency. The latter is due to
the finite extension $l$ of the wave functions in the dots leading to a cutoff of phonons with
wave vectors $Q=|{\bf{Q}}|\gtrsim 1/l$.
The phonons are assumed to be piezoelectric acoustical bulk (three-dimensional) modes 
with an interaction matrix element
$|\lambda_Q|^2=$ ${\lambda^2}/Vc_sQ$,
$\lambda^2=$ ${\hbar P}/{2\rho_M}$,
where $\omega_Q=c_sQ$ is the phonon dispersion, $V$ the volume and
$\rho_M$ the mass density of the crystal, and $P$ the piezoelectric coupling constant. 
Dynamical screening effects of the 2DEG have been absorbed into the value of $P$.
Using typical GaAs parameters \cite{BFS93}, we obtain $g\approx 0.05$.

\begin{figure}[ht]
\unitlength1cm
\begin{picture}(9,7)
\epsfxsize=9cm
\put(-0.5,0.5){\epsfbox{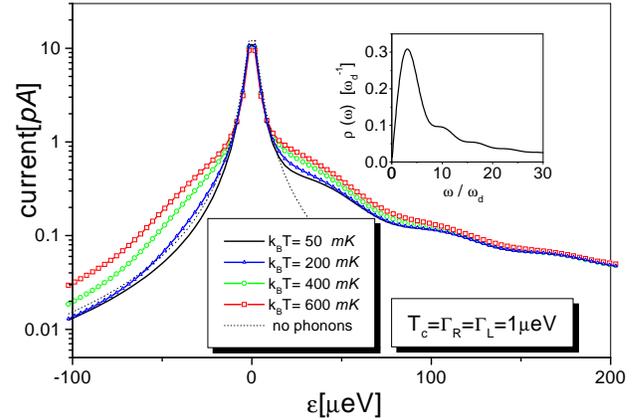}}
\end{picture}
\caption[]{\label{current}Stationary tunnel current, Eq.(\ref{currentstat}), as a function of the energy difference $\varepsilon$ between
left and right dot ground state energies. Dimensionless electron-phonon coupling parameter: $g=0.05$. Inset:
effective density of states $\rho(\omega)$ of phonon modes, Eq.(\ref{oscillator}) and Eq.(\ref{rho2}).}
\end{figure}

The most important feature of $\rho(\omega)$, Eq.(\ref{rho2}),  are 
the oscillations on the  scale $\omega_d=c_s/d$,
(Fig. \ref{current}, inset). In fact, these lead to the oscillations
in the current profile due to phonon emission in the experiment \cite{Fujetal98}:
Fig. \ref{current} shows the stationary current Eq.(\ref{currentstat}) at different temperatures for parameters close to the ones
of the experiment. 
With $d=200\cdot10^{-9}m$ and $c_s=5000 m/s$, we obtain $\hbar\omega_d=16.5\mu eV$, which is
the scale on which the oscillations occur in \cite{Fujetal98}. 
The cutoff energies are assumed as $\hbar \omega_c=1$ meV
and $\hbar\delta=1\mu$eV. A small but
finite value of $\delta$ in $C_{\varepsilon}(\delta\to 0)$ simulates off-diagonal electron-phonon processes which are not
included in $\rho(\omega)$ for $\hbar\omega=\varepsilon\to 0$. 
At low temperatures, there is a broad oscillatory shoulder for $\varepsilon>0$
(spontaneous phonon emission).
It reflects the oscillations in $\rho(\omega)$ which in turn determines
$C_{\varepsilon}$ in Eq. (\ref{currentstat}). Its real part 
$\Re C_{\varepsilon}$  
is proportional to the 
probability density for inelastic tunneling from
the left to the right dot with energy transfer $\varepsilon$ \cite{IN91}.
In the limit $\omega_d=0$, one finds
$
\Re {C}_{\varepsilon}$ $=$ $(2\pi/\Gamma(g))\varepsilon^{g-1}e^{-\varepsilon}\theta(\epsilon)
$ 
at $T=0$, where only spontaneous phonon emission is possible.
This shows that the effect is  non-perturbative  in the
electron-phonon coupling $g$.
As an important consequence, even for  small coupling constants $g$
and  $T\to 0$, the electron-phonon interaction can dominate
the non-linear electron transport. Furthermore, we find an energy dependence
of the current for $\varepsilon\gtrsim 50$ $\mu$eV between $1/\varepsilon$ (larger $g$)
and $1/\varepsilon^2$ (smaller $g$), depending on $g$.
Note that the tunnel current $I$ is determined by the {\em complex} function
$C_{\varepsilon}$ which contains all effects of energy renormalization due to the electron-phonon coupling.
At higher temperatures, the current at the absorption side $\varepsilon<0$
increases  faster than at the emission side where the oscillations start to be smeared out. For
$\varepsilon<0$ a shoulder-like structure appears, consistent with the observation in \cite{Fujetal98}.

We note that our results are  consistent with a previous model \cite{GM88b}
for inelastic tunneling via pairs of impurities. There, perturbation theory in the electron--phonon
deformation potential coupling  was used to obtain the nonlinear current (without oscillations) for large area tunnel
junctions after averaging over an ensemble of impurity pairs.

The spontaneous emission of phonons in double dots is closely analogous to the subradiant spontaneous decay
by emissions of {\em photons} from a laser--trapped two-ion system\cite{DeVB96}. There, the interaction of the atomic
dipoles $\hat{{\bf d}}_i$, at positions ${\bf r}_i$, $i=1,2$  with the electromagnetic field has the form \cite{arg74}
\begin{equation}\label{ion}
H_{eph}=\sum_{{\bf Q}s}{\bf g}_{{\bf Q}s}(\hat{{\bf d}}_1\exp{i{\bf Qr}_1}+
\hat{{\bf d}}_2\exp{i{\bf Qr}_2}),
\end{equation}
with ${\bf g}_{{\bf Q}s}=$ $-i(2\pi cQ/V)^{1/2}\vec{\varepsilon}_{{\bf Q},s}$ with
the speed of light $c$, and polarization vectors $\vec{\varepsilon}_{{\bf Q},s}$ for polarization direction $s$.
This leads to rates for super- and subradiant decay of the form \cite{Benedict}
$\Gamma(Q)_{\pm}=\Gamma_0(Q)[1\pm\alpha\sin(Qd)/(Qd)]$ with $\alpha = 1$ ($\alpha=3/2$)
if the vector character of the light is (not)
neglected, $Q=\omega/c$, and $\Gamma_0(Q)\propto Q^3$. The subradiant channel ($-$) is due to the decay
of the singlet state that corresponds to the {\em difference} $\hat{{\bf d}}_1\exp{i{\bf Qr}_1}-
\hat{\bf {d}}_2\exp{i{\bf Qr}_2}$ of the dipole moments. The interaction with a {\em phonon} of mode ${\bf Q}$  in the double dot is
$\propto n_L\exp{i{\bf Qr}_1}+n_R\exp{i{\bf Qr}_2}$. Thus, it has the same `interference form' as in the
two-ion  case, Eq.(\ref{ion}).
The tunnel current is modified by the phase {\em difference} of the electron before and after tunneling whence $\rho(\omega)$,
Eq.(\ref{rho2}), corresponds to the subradiant rate $\Gamma_-$. Although the microscopic mechanism is different in
both cases (for light the rates $\Gamma_0$ are $\propto \omega^3$, for piezoelectric phonons $\propto 1/\omega$),
the interference term $\sin(Qd)/Qd$ in both cases is due to the `interference of matrix elements' in the
interaction Hamiltonian and  has the same physical origin.

In the atom--trap experiment, the Dicke effect, i.e. the existence of two different radiation channels $\Gamma_{\pm}$, has been
verified by changing the ion distance. The experimental data of the double quantum dot \cite{Fujetal98} seem to indicate that for larger
distance $d$ of the dots the oscillations $\propto \sin(\varepsilon/\hbar\omega_d)$ as a function of $\varepsilon$
become faster which is consistent with $\omega_d=c_s/d$ in Eq.(\ref{rho2}).

In conclusion, we have found an interference effect which explains the recently observed
phonon emission spectrum in the transport through coupled quantum dots. 
The analogy between artificial and real multiple atom systems allows to speculate about further realizations
of coherent optical effects in semiconductor quantum dot experiments, similar to
the oscillatory Dicke effect we have found recently \cite{BIS98}. In the case of phonons discussed here,
the coherence (oscillations due to interference) showed up {\em within} a dissipative process, i.e. spontaneous
emission. In order to further study such `coherent dissipative phenomena', we suggest
systems of two of more double quantum dots which are coupled via the common phonon field.
The non--linear transport properties
are then determined by the electron--phonon coupling both within each and between the double dots.

This work has been supported by
the TMR network
`Quantum transport in the frequency and time domains',
the Graduiertenkolleg `nanostrukturierte Festk\"orper' (Hamburg), the
SFB 508 `Quantenmaterialien' (Hamburg), and the DFG project Kr 627/9-1.

\end{document}